# Exploring High Performance Distributed File Storage Using LDPC Codes

Benjamin Gaidioz
CERN
benjamin.gaidioz@cern.ch

Birger Koblitz
CERN
birger.koblitz@cern.ch

Nuno Santos
CERN and Univ. of Coimbra, Portugal
nuno.santos@cern.ch

September 25, 2018

## Abstract

We explore the feasibility of implementing a reliable, high performance, distributed storage system on a commodity computing cluster. Files are distributed across storage nodes using erasure coding with small Low-Density Parity-Check (LDPC) codes which provide high reliability while keeping the storage and performance overhead small. We present performance measurements done on a prototype system comprising 50 nodes which are self organised using a peer-to-peer overlay.

## 1 Introduction

With the growing popularity of Grid and cluster computing, computer clusters built out of cheap commodity computers are becoming commonplace. While their CPU power is readily made available through batch or grid computing systems, the often substantial amount of disk space on the computer nodes is usually not made available for mid or long term storage. In this paper we investigate how to make this space available to be used as high performant and reliable file storage for applications where files are written once and read often.

Statistical analysis on the availability of distributed files shows that erasure coding, where the file is decomposed into $n$ *data* and $m$ *coding* blocks of equal size and can be reconstructed from any $n$ blocks. Statistical analysis shows [11, 19] that when the hosts on which the data is stored are relatively reliable, erasure coding is able to ensure a much higher availability than full file replication, while introducing a smaller storage overhead. However, in practice, most of the storage systems used on Local Area Networks rely on replication to ensure reliability. The reason is that traditional erasure coding like Reed-Solomon-Codes demand a high computational effort, which grows quadratically with $n$ and $m$, to reassemble the original data out of any $n$ data or coding blocks.

Low-Density Parity-Check codes (LDPC) [6] provide a solution to this problem because they allow to reconstruct the original data using relatively few and cheap XOR operations. They do not, however, code the data optimally (in contrast to Reed-Solomon codes) but require $fn$ blocks to reconstruct the stored file, where $f \geq 1$. The properties of LDPC codes are well understood in the asymptotics of $n \to \infty$ where $f \to 1$, but little is known about how to construct smaller codes ($n, m < 1000$). The discovery of efficient algorithms for creating large LDPC codes ($n > 10000$) with very fast encoding and decoding [12] has lead to a surge in the interest in these codes, in particular for the resilient storage of files on Grid and peer-to-peer networks. In this scenario a file is decomposed into many ($n+m$ large) blocks which are stored in a distributed manner on hosts connected by a Wide Area Network (WAN) [3].

The paper is organised as follows. In section 2, we compare the availability provided by LDPC codes versus erasure coding and replication. We explain why small LDPC codes ($n, m \approx 10$) fit better the criteria needed for the implementation of a distributed storage system in a commodity computing cluster. These small codes cannot be constructed with standard techniques. We therefore present a way of constructing graphs with good guarantees on their redundancy using Monte Carlo techniques in section 3. In section 4, we describe the implementation of a file storage system based on small LDPC codes. Performance measurements are presented in section 5. The remainder of the paper is a discussion of the implementation and the results obtained so far including references to related work (section 6) and finally conclusions and a preview of our ongoing and future work (section 7).

# 2 Availability Analysis of LDPC vs Erasure Coding vs Replication

In the following we will give an overview on the storage overhead and availability of normal erasure codes, LDPC codes and common file replication. A detailed study can be found in e.g. [11, 20].

Written in a common form, the availability of a file replicated $S$ times ($S = (n+m)/n$ is called the stretch factor) are

$$A_r(\mu, S) = \sum_{i=1}^{S} \binom{S}{i} \mu^1 (1-\mu)^{S-i} \qquad (1)$$

and for an erasure coded file

$$A_e(\mu, n, m) = \sum_{i=n}^{n+m} \binom{n+m}{i} \mu^i (1-\mu)^{n+m-i} \quad , \qquad (2)$$

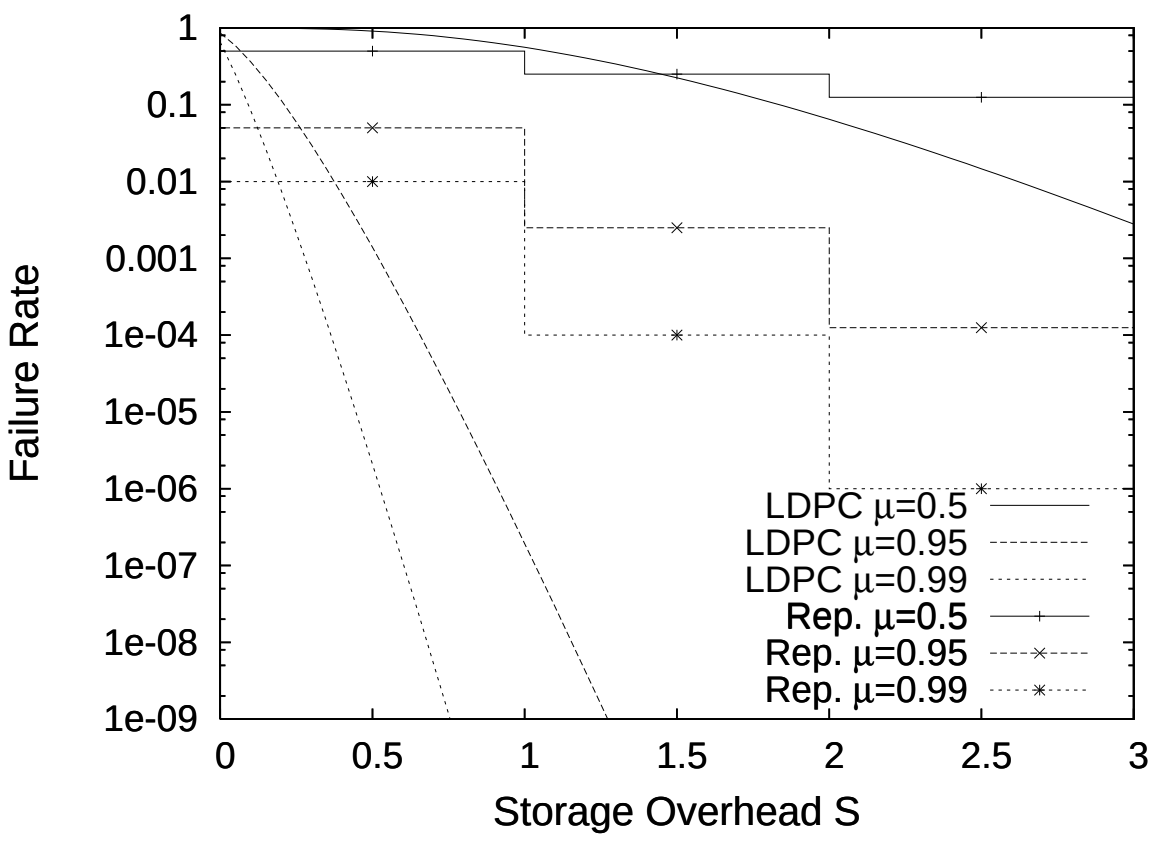

Figure 1: The average failure rate of files stored redundantly using replication and using LDPC codes versus a given storage overhead. We choose $n = 8$ for the LDPC codes and assume an overhead factor of $f = 1.1$.

where $\mu$ is the availability of a host. The rate of the codes is

$$R = \frac{1}{S} = \frac{n}{n+m} \quad . \qquad (3)$$

LDPC codes which do not code optimally introduce an overhead factor $f$, this means they are only able to reconstruct the original file from in average $fn$ chunks, where $f > 1$. Concerning the overhead, LDPC codes are comparable to normal erasure codes with

$$n' = fn \quad \text{and} \quad m' = (1-f)n + m \quad . \qquad (4)$$

An upper bound for the availability of an LDPC encoded file can be given by (2) using $f_{\max}$, the maximum overhead of a graph (that is the original data can be reconstructed from *any* $f_{\max}n$ blocks).

Figure 1 shows a comparison of the failure rate $(1-A)$ of files stored redundantly using replication with LDPC encoded files with $n = 8$ and with an assumed overhead factor of $f = 1.1$ for three different node availabilities of $\mu = 0.5$, 0.95 and 0.99. For bad availability of files, LDPC codes with such small number of coding blocks perform worse than file replication, at least for small storage overhead factors of $S < 1.5$. However for a good availability $\mu = 0.95$ (an estimate for the availability of nodes in a cluster of custom hardware) or even 0.99 small LDPC codes can provide a much better availability of the files for smaller overheads. This means that small LDPC codes have the potential to provide small storage overhead and excellent file availability on LANs while introducing only a small networking overhead due to the relatively small number of $n$ parallel downloads.

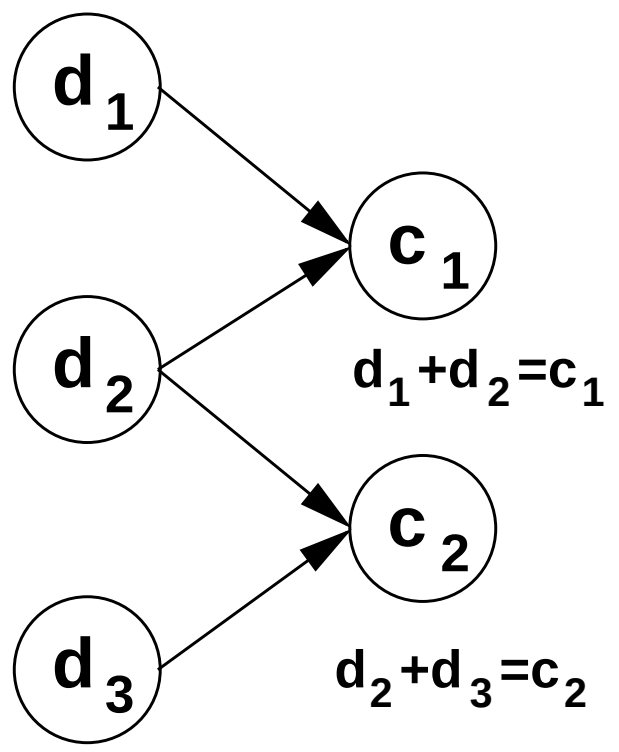

Figure 2: An example of a systematic graph with $n = 3$ and $m = 2$.

# 3 LDPC Codes

In the following we will give a brief overview of LDPC codes, for a full introduction see e.g. [21]. An example of a graph describing a simple LDPC code is shown in fig. 2: From $n = 3$ data words $d_1, d_2, d_3$ (bits in the simplest case), $m = 2$ coding words $c_1, c_2$ are calculated by *xor*ing the coding words. For example $c_1 = d_1 + d_2$. Encoding the redundant information in the coding words is done in a time growing linear with the number of edges in the graph. The data and coding words are then assembled into data and coding blocks and can be stored in a distributed manner.

The original information of a file can be reconstructed directly from the data blocks by simply concatenating them. This will be possible in the majority of cases if they were distributed locally on a relatively reliable LAN. If data blocks are unavailable then they can be reconstructed from coding blocks using the following algorithm: If for a known coding block all but one of the data blocks from which it has been calculated are known, then the words in that unknown data block are the exclusive or of the corresponding words in the coding block and the known data blocks. By applying this algorithm recursively to the downloaded or reconstructed blocks, the original data can be reconstructed in linear time, if a sufficient number of data and coding blocks is available.

The amount of information encoded using an LDPC graph is the rate $R = n/(n+m)$. Since LDPC codes do not encode optimally more than $n$ blocks are needed to reconstruct the original file, when randomly downloading blocks. The overhead factor $f$ is defined by the average number $fn$ of blocks which need to be downloaded to reconstruct the file, where $f > 1$. In the limit of very large $(n, m \to \infty)$, LDPC codes become optimal $(f \to 1)$, for small and medium sized codes $(n, m < 10000)$ the overhead is typically in the order of 10%, depending on $R$ [14].

When using large LDPC codes for the distributed storage on WANs, codes with $f$ as small as possible are needed which allows to pick blocks for download based on latency or available bandwidth. However, for the usage we envision, that is the reliable and high-performance storage of files on a LAN, a small $f$ is not necessarily required since the availability of the blocks will be good and the available bandwidth does not differ for the distinct blocks (ignoring for now the problem of "hot" files and blocks). In fact for performance reasons in the normal case a client will try to download only the $n$ data blocks in order to be able to reconstruct the original file by simple concatenation. However, we are also interested to give certain guarantees on the availability of a file, which means we need to know the worst-case overhead $f_{\max}$, such that we can guarantee a successful download of a file in case of $f_{\max}n$ blocks of the file being available.

## 3.1 Generating Efficient Codes

For high performance and reliable storage of files on a LAN, intended to replace a typical disk-server with several ordinary computer nodes, optimal codes of relatively small ($n, m = \mathcal{O}(10)$) size are necessary. This scale is set on the one side by the fact that such disk servers normally have about ten times faster network links and hard-disks which can be achieved in a distributed storage by downloading several coded blocks in parallel, on the other side by the fact that large $n + m$ introduce an overhead on the network as well as for the organisation of the system. While the construction of very small ($n \in \{2, 3\}$ and $m \in \{3, 4, 5\}$) optimal codes has been recently done [13], larger codes can only be constructed and evaluated by Monte Carlo techniques [15].

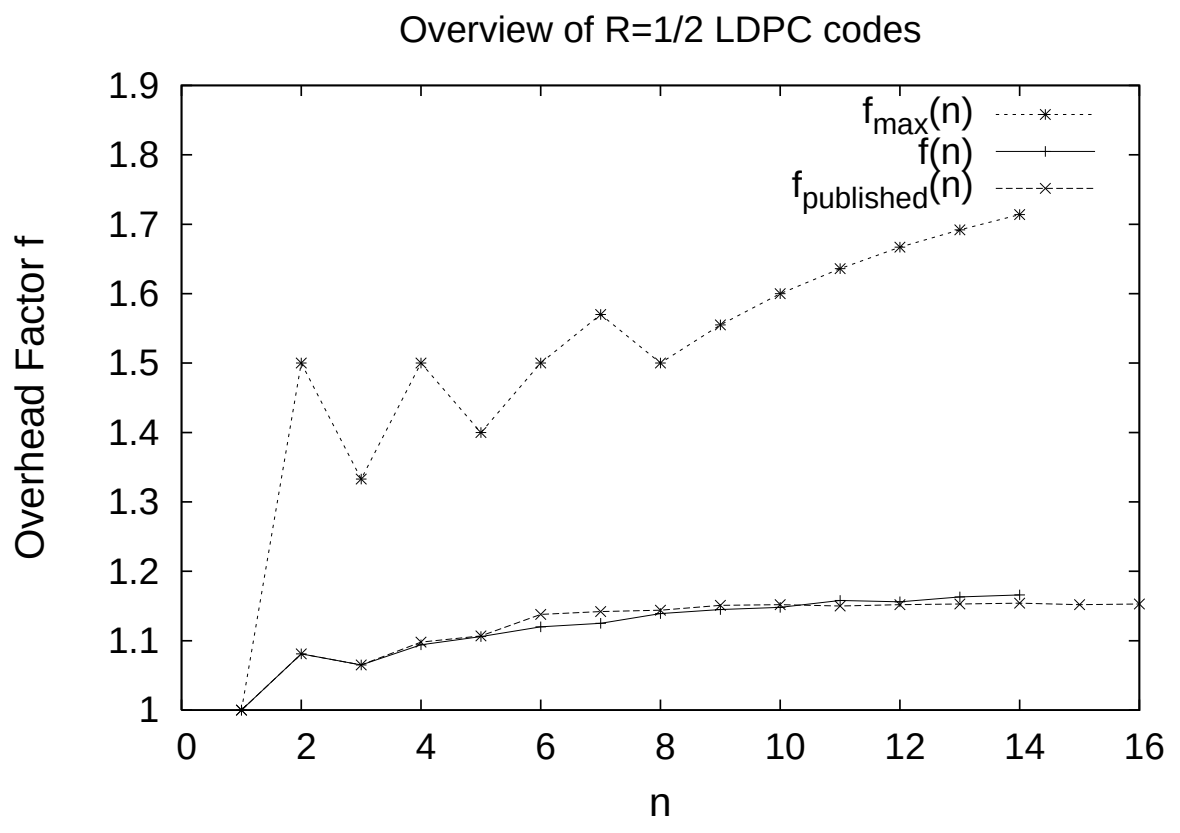

Figure 3: Average overhead $f(n)$ of best generated codes compared to the overhead of published codes $f_{\text{published}}(n)$ (taken from [15, 13]) for $R = 1/2$. In addition the worst-case overhead factor $f_{\max}$ of the best generated graphs is shown.

Using such a Monte Carlo technique we create graphs randomly for a fixed $n$ and $m$ and a probability $p$ for a right-hand node to be connected to a given left-hand node. We use $0.4 < p < 0.6$ depending on $n, m$ based on the findings in [13]. Instead of evaluating the average overhead factor by sampling the necessary overhead for many different downloading sequences of blocks, we compute $f_{\max}$ for the given graph. Calculating the average overhead with $s$ samples in average

$$E = s \prod_{i=n}^{fn} i \tag{5}$$

evaluations whether with the given set of blocks the original data can be reconstructed. The resulting error on $f$ is $f/\sqrt{s}$. Computing $f_{\max}$ requires at most

$$E' = \prod_{i=f_{\max}n}^{n+m-1} i \tag{6}$$

reconstruction tries where $E' < E$ for small graphs ($n \ll s$). Since in practice most generated graphs will be unable to cope with even a very small number of missing coding blocks and $f_{\max}$ is found as soon as a single download sequence fails, the average number of tries is close to $n(n-1)$ (most graphs failing to compensate for 2 missing data blocks in all cases). Figure 3 shows that the performance of the graphs generated and evaluated in this manner for $R = 1/2$ can compete with the best known graphs for $n < 15$.

## 3.2 Performance of LDPC Codes

In this section, we evaluate the performance of LDPC codes of different rates in order to select good values for $m$ and $n$. We then use the solution presented in section 3.1 to generate a graph with good properties. We implement this graph in our storage system prototype and evaluate the overhead of decoding with a varying number of missing data chunks.

Fig. 4 shows the system failure rate provided by four different code rates as a function of the number of data chunks $n$. Rates like 1/2 and 1/3 provide a low failure rate at the price of a high storage overhead. Rate $R = 2/3$ has a low overhead but a high failure rate. For example, with $R = 2/3$, we need $n = 14$ and $m = 7$ for reaching a failure rate of $10^{-6}$, while the same availability can be obtained with a $R = 4/7$ (which has almost the same storage overhead) with $n = 8$ and $m = 6$. For our availability goals, rate $R = 4/7$ provides a good tradeoff in storage overhead and availability. We use $n = 8$ and $m = 6$.

To generate a good graph for the parameters $n = 8$ and $m = 6$, we ran the graph generation algorithm presented in section 3.1. The resulting graph is shown on fig. 5. It has the property of tolerating the loss of any three data chunks, that is, the file can always be reconstructed even if any three data chunks are missing.

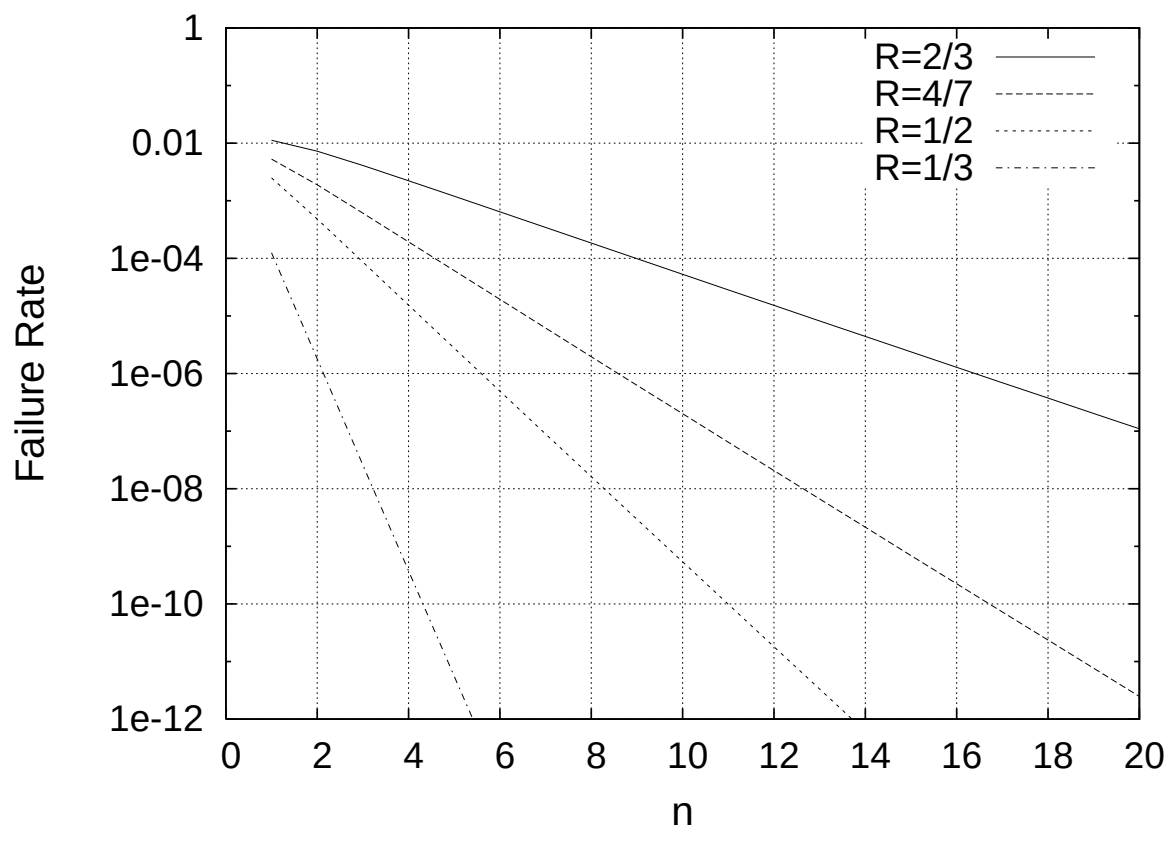

Figure 4: Failure Rate as a function of the number $n$ of data chunks for different code rates. We assume a node availability of $\mu = 0.95$.

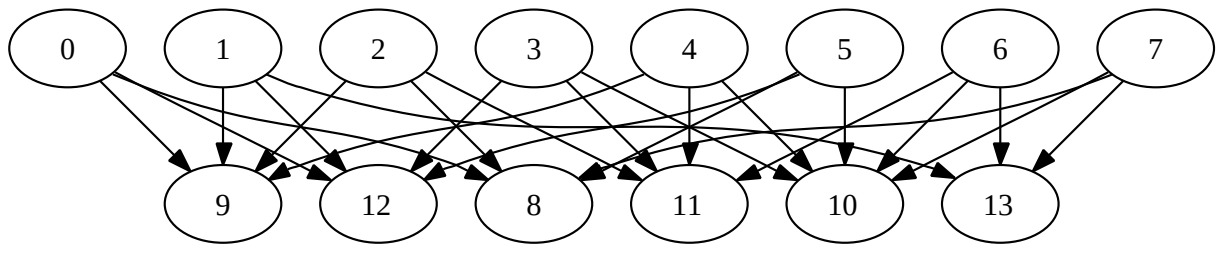

Figure 5: An example of a graph with $n = 8$ and $m = 6$ which can be reconstructed with any 11 nodes. The average overhead factor is $f = 1.108$.

Reconstructing a file out of both control and data chunks has some overhead because it requires computations instead of simple concatenation of chunks. We evaluated this overhead by decoding files with variable number of missing chunks. The tests were run on the Münster testbed. Tab. 1 shows the data rates obtained when downloading chunks from several servers, as a function of the number of failing data chunks. The file size is 500 MB. It was encoded using the graph shown on fig. 5. Since this graph allows to reconstruct a file with up to three missing chunks out of fourteen, we vary the number of missing data chunks from zero to three. The client also ran on the Münster cluster (see sect. 5 for a description of the cluster). We show here the average of 50 downloads.

Table 1: Performance of reading a 500 MB file as a function of missing data chunks.

| missing chunks | time (s) | rate (MB/s) |
|---|---|---|
| 0 | 4.79 | 109.5 |
| 1 | 5.23 | 100.25 |
| 2 | 5.73 | 91.5 |
| 3 | 6.47 | 81.0 |

There is an obvious cost in decoding missing data chunks out of control chunks. However, the performance is still acceptable, even in the worse case of three missing chunks. Considering that the worse case will not occur often in a LAN, the overhead in general would be much smaller.

# 4 A Prototype Implementation

A prototype for the system presented in the article has been implemented. We give in this section a description of the implementation.

## 4.1 Overall Architecture of the System

Files will be distributed to servers running on nodes involved in the system. Applications using the system are linked to a client library.

**Servers** The system is deployed over a set of nodes in a computing center. Each host runs an instance of the *server*. The server is responsible to host file chunks, distributing them to clients (and also to receive them when a client stores a file in the system). We use HTTP both for file transfers and for control messages.

We associate with each server a unique identifier obtained by computing a hash on the name of the host it is running on. This implicitly defines an order on servers. When looking for a file chunk (either for storing it or retrieving it), the same hash function is computed on the chunk name and the server having the closest hash value is identified as the one hosting it.

**Clients** Access to files is implemented in a C++ client library that applications link against. The library currently provides one call to make a local copy of a stored file or to copy a local file to the system. Files are identified by a flat file name. We do not intend to implement a file system interface to the system but rather a "file server" interface.

When a client is started, it needs to update its local list of hosts in the system so that it will be able to get file chunks from them. For bootstrapping, clients are configured manually with a list of well-known hosts that they contact at startup to obtain an updated list. Clients store their list persistently, so that when they are restarted they can quickly reestablish contact with existing hosts. The list of hosts is kept consistent using a peer-to-peer overlay network. See sect. 4.4 for more details.

## 4.2 Storing a File

File chunks are randomly stored on a set of hosts. To store a file, the client first splits it into $n+m$ chunks and assigns to each a name consisting on original filename plus the index of the chunk. It then computes the hash of each chunk's name. Each chunk is stored on the server whose identifier is closer to the hash. Using a hash function enables chunks to be randomly distributed in the system. To read a file, the client can contact the hosts storing the chunks directly by just knowing the chunk names.

In case of failure of some destination nodes, the current implementation discards the failing chunks, not storing them on the system. If the missing chunks are less than three, clients will still be able to read the file from the available ones. This is of course not the ideal solution and we are working on mechanisms for tolerating these failures. This question is addressed in sect. 4.5.

## 4.3 Retrieving a file

In order to download a file, the client computes the hash of each chunk's name to determine the hosts storing each chunk. Initially, it tries to download only the $n$ data chunks. Unless some nodes are unreachable or some chunks are not found, these $n$ fragments are sufficient to reassemble the file very efficiently, as they only have to be concatenated. In case of failure, the client uses the graph to find the control chunks required and downloads them instead. It is possible that some control chunks fail, in which case the graph may permit to rebuild them with other chunks. If the available chunks do not permit to reconstruct the file, that is, if there are more than three missing chunks, the call fails.

HTTP provides a way of obtaining a slice of a file by specifying a start and an end offset. We use this to switch from a failing server to another in the middle of a file transfer. However, the client API does not yet provide a call for downloading a specific *part* of a file. The file is download entirely. This feature would permit the client library to ignore chunks which would not be needed (in case the data length allows it).

There are a couple of interesting strategies to implement in downloading a file in a robust manner. In the current implementation, we have only implemented recovery based on the graph. However, other optimisations for improving the availability of files, with strategies implemented in both clients and servers can be envisaged and are presented in sect. 4.5.

## 4.4 Consistency of the host list

Having a client contact the right servers for getting the right file chunks strongly relies on it initially obtaining a consistent list of hosts. We ensure consistency of the host list using lightweight group communication between servers. When a client starts, it asks any server for this list and uses it to download chunks.

We use weak-consistency for maintaining the host list. Although strong consistency would be desirable, we consider that it is more important to provide good scalability and accept some temporary inconsistencies. The protocol is designed in a way that if the system is left on a stable state, the information on the hosts will converge (eventual consistency).

An interesting feature is that although the weak consistency means that a client can start with a wrong host list, it does not necessarily imply the service becomes unavailable. After the system is running for some time, there might still be in the system some hosts which have a slightly inconsistent host list and would then provide a slightly wrong information to clients. A client receiving the host list from one of these nodes would then fail to download some chunks, not because of node failure but simply because it contacts the wrong nodes. In this case, it would still implement recovery procedures and hopefully successfully download control chunks in order to rebuild the file.

In a similar way the whole system may still provide 100% availability with some chunks permanently lost. The same applies in the case the host list clients start with is slightly inconsistent.

Host list consistency is implemented using a rumour mongering protocol [8] which combines pushing and pulling of rumours. At random times, a server sends (pushes) to a subset of its known peers information like "host $H_1$ joined", "host $H_2$ left". The information is then forwarded to further nodes with a decaying probability. Additional information on host availability is gained by the rumour mongering process itself (hosts contacting each other are in fact pulling information on their availability).

Assuming the frequency at which nodes enter and leave the system is large compared to the rate at which information is "gossipped" in the system, the information converges to a stable and consistent state. Convergence speed is important when the system is started and can be improved by tuning this frequency. However, after the system has been running for a long period, inconsistencies should be harmless as explained above and a stable service should be available.

## 4.5 Fault tolerance

We have explained how the system achieves robustness by using LDPC codes for encoding files. There are several interesting optimisations to be put in place concerning fault tolerance which are explained in this section.

**Reconstruction of lost blocks** Servers can decide to host missing chunks due to another host failure. In case a server disappears, all the chunks it is hosting are lost. Due to the erasure coding, this does not necessarily entail loss of files. However, for improved reliability, the host which is closest in the host hash ring could recreate the chunks from available ones so that they would be found again. This should not be implemented systematically. However, once the host list has definitely lost the host and this server constantly gets polled for these lost chunks, it would take the initiative of hosting them.

**Client's host list** In the current implementation, the client library gets a host list from the first available node and uses it to store files. This may be an inefficient strategy when the system is bootstrapping and many nodes have an incorrect host list. We are implementing checks in the client library by getting several lists simultaneously in order to increase the validity of the initial host list.

**Uploading failing chunks to alternative hosts** We have explained above that the process of writing files is implemented in a rather optimistic way: if a destination host is failing, the chunk is simply not stored. Unless the client would get a more up to date list of host and store the chunks normally, a strategy could consist of storing the chunks in hosts which would take over from failing servers. This would increase availability without waiting for these servers to reconstruct the missing chunks themselves.

**Load balancing** When a server is overloaded it could refuse the download requests from some clients. These clients could then contact another server to download an alternative control chunk. The client would need to decode the file from the control chunk or even have to download more than $n$ chunks, but this might still be faster if the transfers can complete more quickly because the load is better balanced. Implementation of this load balancing strategy requires a very efficient implementation of decoding, so that clients can still reconstruct the file quickly.

### 4.6 File server operations

We have chosen to provide access to the system in a similar way to a file server with broad file access and a simple semantic. Features mandatory to file access are not the topic of this paper which concentrates on the use of LDPC codes for storing files.

**File inventory** Currently the system itself does not know which files are stored. Instead we rely on an external catalogue to keep an inventory of the stored files.

**Access control** The software here does not include sophisticated file access (access control, file updates, etc.). We intend to use this system in production and this type of feature will be quickly implemented.

## 5 Performance

In this section, we show performance measurements obtained with our prototype implementation. The main goal of these measurements carried over these testbeds is to have an idea of the overall data rate one can achieve with a system like the one presented here. Deeper studies regarding availability are ongoing work which we intend to implement through simulations instead. This would permit to evaluate the robustness of the system with various failure pattern. Later, we shall run the system in a production like manner.

We present the testbeds in sect. 5.1 and measurements are shown in sect. 5.2. LDPC codes are known to be efficient in terms of decoding. We measure the overhead of downloading a file with more or less reconstruction involved in an other section dedicated to performance of LDPC codes (sect. 3.2).

### 5.1 Testbeds

We ran the prototype on two different platforms.

**CERN "lxplus" computing farm** is a cluster of about 100 dual Xeon 2.8 GHz with 2 GB of RAM with a Fast-Ethernet access to the network. We used 40 of them.

**Münster cluster** is a set of 50 dual Opteron 2 GHz with 2 GB of RAM. There are interconnected with a Gigabit network.

On each node, we run a server. Then we store files to the system by sending chunks to servers. When the files have been stored, we start a client on each of the nodes.

Clients are configured to download all the files in a random order so that at any time, they access different files. They also start downloading files one after each other so that we have measurements for a different number of clients running in parallel.

Both clusters are used in production and many nodes are very busy with computations during our tests. This is not a drawback because our system is intended to be under such conditions in practice.

### 5.2 Measurements

Measurements are shown of fig. 6 and 7. We plot on fig. 6 the aggregated rate of the whole system as a function of the number of clients running. On fig. 7, we plot the equivalent data rate it provides per client.

**Aggregated data rate** Overall rate measured on both clusters is plotted on fig. 6. We see that in both cases, the rate increases each time a client enters in the system. For a number of clients larger than 10 to 12 nodes, the overall rate stops growing. On lxplus, the maximum rate is about 110 MB/s. It is 350 MB/s in the Münster cluster.

**Rate per client** It is shown on fig. 7. Rate per node obtained on lxplus remains quite constant for a number of clients between 1 and 12. It is about 10 MB/s. For a high number of clients, the rate decreases to 3 MB/s. On the Münster cluster, the rate per node decreases from 80 MB/s to 7.15 MB/s for 50 clients.

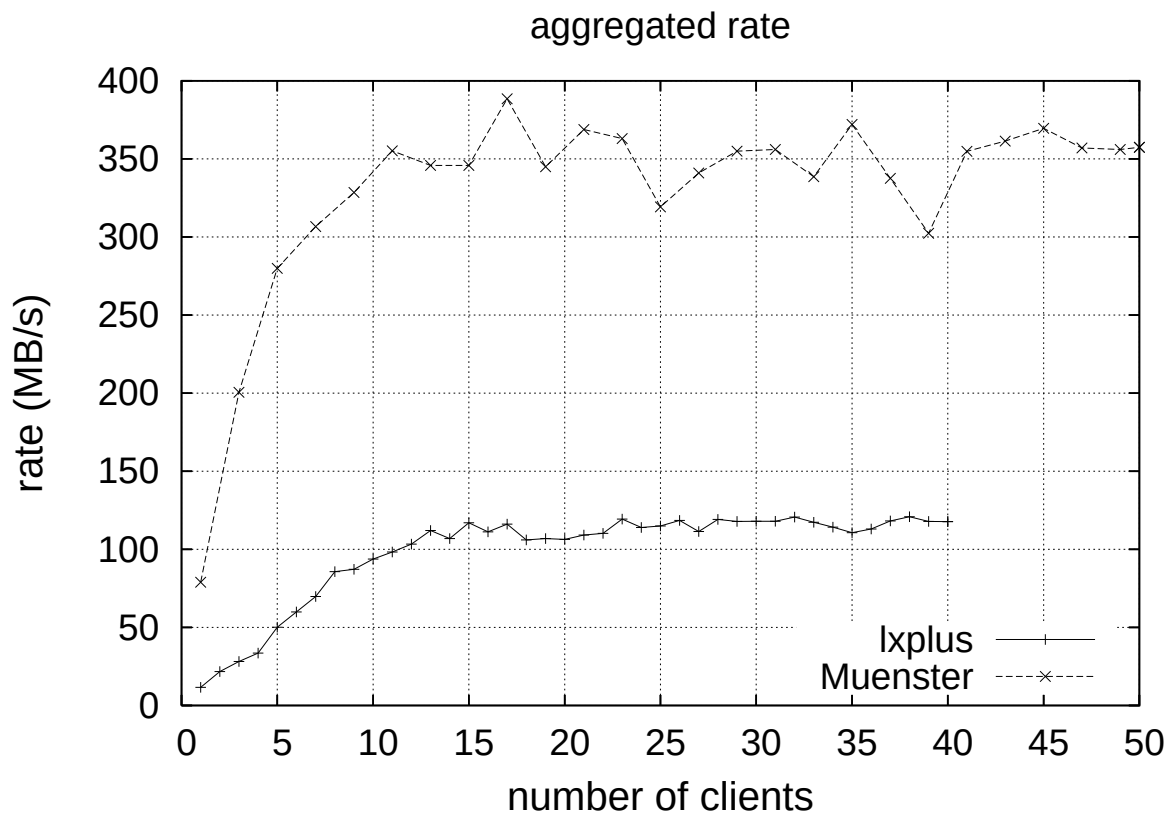

Figure 6: Aggregated rate obtained on both clusters as a function of the number of client nodes.

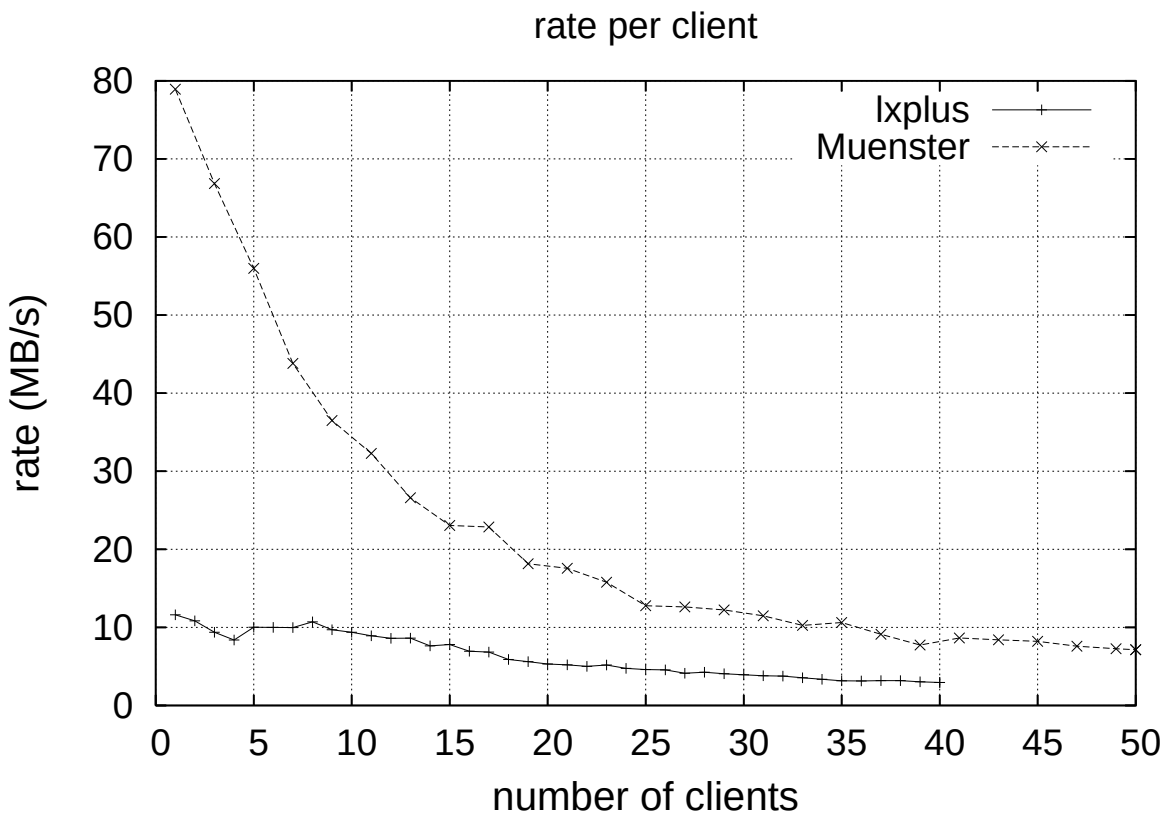

Figure 7: Rate per node obtained on both clusters as a function of the number of client nodes.

For a large number of hosts, the whole set of servers and clients obviously hits a limit of the cluster which depends on the average CPU, disk and network usage.

For a low number of hosts, although the rate per node decreases when increasing the number of clients, the loss in performance is relatively low compared to the number of nodes. On lxplus, the gain in performance per node is about 9.78 MB/s. On Münster's cluster, the increase in performance per node is about 50 MB/s. While performing these measurements, we noticed the increase in rate is lower than what one would expect. We know that TCP transfers can run at a higher rate on this type of hardware.

We believe this is a consequence of some nodes being more loaded than others and the fact we keep connections synchronised for avoiding storing chunks on disk. When ran against CPU intensive tasks running in the same system, the performance of both sending and receiving data drops significantly. For example when running tests on Münster's cluster, we noticed a simple file transfer can run at 12 to 20 MB/s from certain busy nodes. Running systematic tests with several TCP transfers in parallel, we have observed that these connections not only suffer from running on a busy node. The fact we are keeping chunk downloads synchronised (so that decoding is performed in memory) leads the well performing connections to send traffic in bursts when a block has been decoded. Such pattern affects even more the already low speed of slow connections because they get interrupted often and system timeouts lead them to restart in slow-start more often. The rate of a whole file download is affected by this as we could observe it in system tests while preparing these experiments.

We are investigating the problem but we consider the numbers shown in this section promising. In fact, since we intend to implement this storage system on commodity computing clusters, we expect data rate to be affected more strongly by other factors like other tasks running on the nodes, the disk usage, and network usage. The most critical goal is to ensure high availability of the service.

# 6 Related Work

Reed-Solomon codes have been used by several storage systems, both for WAN environments

(OceanStore [16]) and for LAN (RepStore [10] and FAB [5]). All these systems use erasure coding only for archival storage, since Reed-Solomon codes have a significant storage overhead. For frequently accessed files and for supporting updates, these systems rely on replication. Many other distributed storage systems rely solely on replication, including Gnutella [17], CFS [4] and PAST [18] for WANs, and Petal [9] and the Google File System (GFS) [7] for LANs. In contrast to these systems, we rely solely on erasure coding by using LDPC codes. This is possible due to the LDPC codes' near real-time decoding speed, which allows us to have space efficiency without sacrificing performance.

LDPC codes have not been explored as much for storage applications. The most significant example of their use is the Digital Fountain system [2], where LDPC codes are used for the dissemination of bulk data to a large number of receivers over Wide Area Networks. There is also some recent work [3] that studies the suitability of LDPC codes for Wide Area Network storage. But to our knowledge, there is no previous work on the use of LDPC codes for storage on a Local Area Network environment, where the focus is on performance and a small storage overhead.

Our system is based on a Peer-to-Peer topology due to its fault-tolerance and scalability properties. For the same reasons, many other storage systems use Peer-to-Peer or decentralised topologies. On the Wide-Area Network some examples include Gnutella, CFS, PAST and OceanStore. Gnutella is based on an unstructured topology, while the other three systems are structured using a distributed hash table. For Local Area Networks, xFS [1], Petal and FAB are all decentralised systems that rely on voting and consensus algorithms for organising their topology. Also for the Local Area Network, RepStore uses a one-hop distributed hash table. In contrast to these systems, our gossiping protocol is more light weight and scalable, having as a drawback the possibility of temporary inconsistencies.

# 7 Conclusion and Future Work

We have presented a novel architecture for a reliable, high performance, distributed storage system on a commodity computing cluster. Storage of files is based on erasure coding with small Low-Density Parity-Check (LDPC) codes. These codes provide high reliability given a low storage and performance overhead. The main contributions of this paper are:

- an analytic evaluation of the availability provided by LDPC codes versus replication and erasure coding,
- a way of constructing small LDPC codes with good guarantees on their redundancy,
- the description of an implementation of a file storage system based on LDPC encoding and performance measurements obtained with it on two different computing clusters of both the overall rate it provides and evaluation of the overhead of decoding.

Availability provided by LDPC encoding techniques makes it a satisfying redundancy schema for the implementation of a storage system on a computing cluster. Our work on generation small graphs allows us to obtain a good availability of the service against possible failures of nodes. The initial performance results are promising.

The work presented here is ongoing work and many interesting details are under study.

- Techniques regarding LDPC codes are still being investigated. We continue our activity on generating good graphs with more sophisticated ways of controlling the probability distribution of the edges in the graphs as proposed in [14].
- So far we presented an analytic evaluation of the availability provided by the use of LDPC codes. In the future we intend to use simulations of the entire system instead, so that we can study various failure patterns, e.g. introduced due to failures in the peer-to-peer overlay.

- The implementation itself is still at an early stage. We mainly provide a recovery based on the underlying LDPC graph. Using the peer-to-peer overlay it would be also possible for the system to actively recover missing blocks. In fact, given the nature of the coding graphs this would involve only a small number of hosts which have blocks that are related to the missing one.

- There is currently no load-balancing done by the implementation apart from the trivial case that node becomes unavailable due to their load such that blocks are taken from elsewhere. However, the servers could also distribute information about their load through the P2P network and actively reroute clients or initiate further replication.

- Since we intend to use this system in production, in particular on grid computing sites, part of our activity will be dedicated to its integration into grid file catalogues which will also allow to implement access controls.

# 8 Acknowledgements

This work was performed within the LCG-ARDA project and the authors would like to thank in particular Massimo Lamanna (CERN) for many fruitful discussions on this paper's subject. For the performance measurements a computer cluster at the University of Münster, Germany was used and we are very grateful for the excellent operator support we got. This work was partially funded by grant SFRH/BD/17276/2004 of the Portuguese Foundation for Science and Technology (FCT) and by Bundesministerium für Bildung und Forschung, Berlin, Germany.

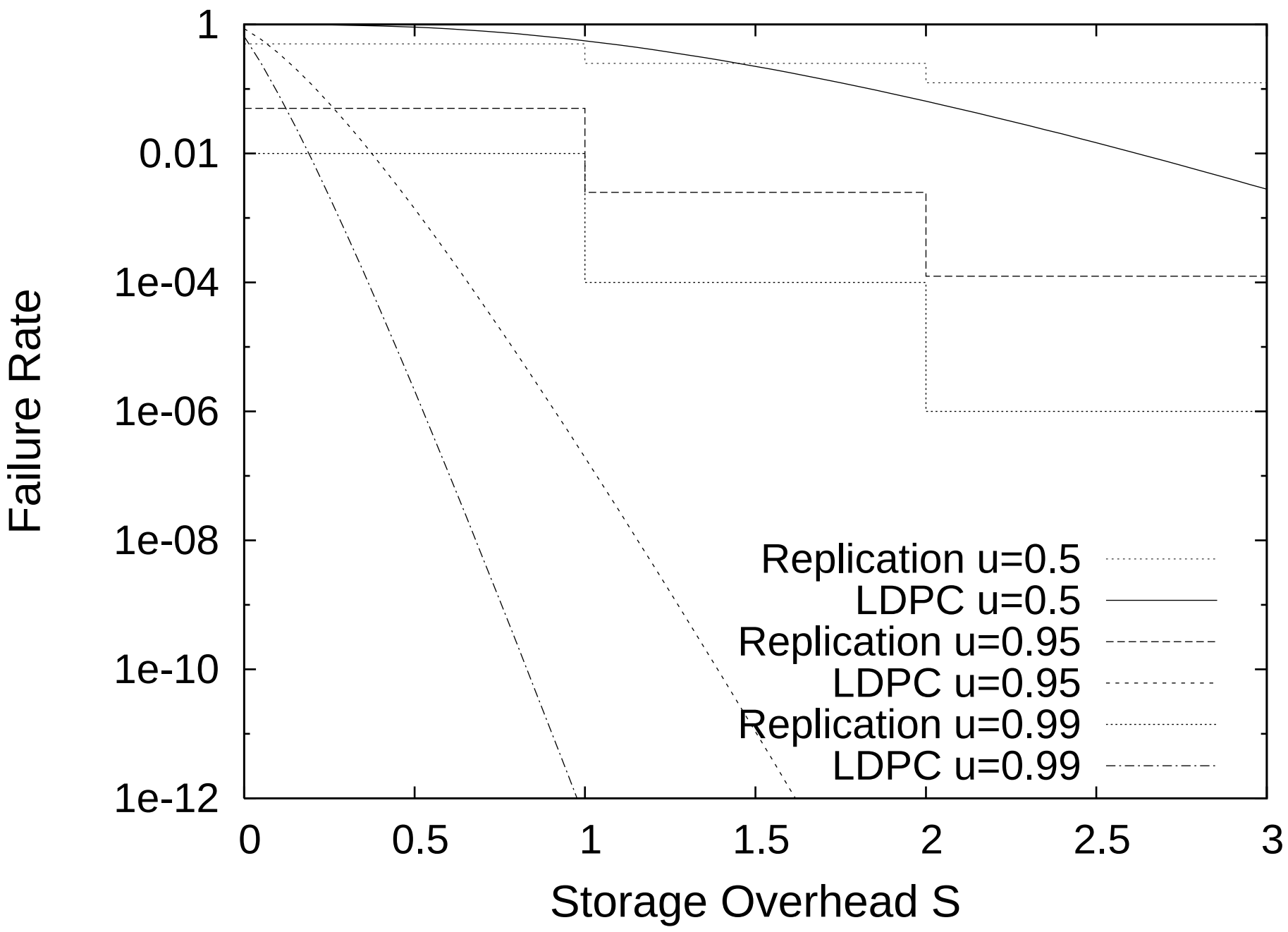

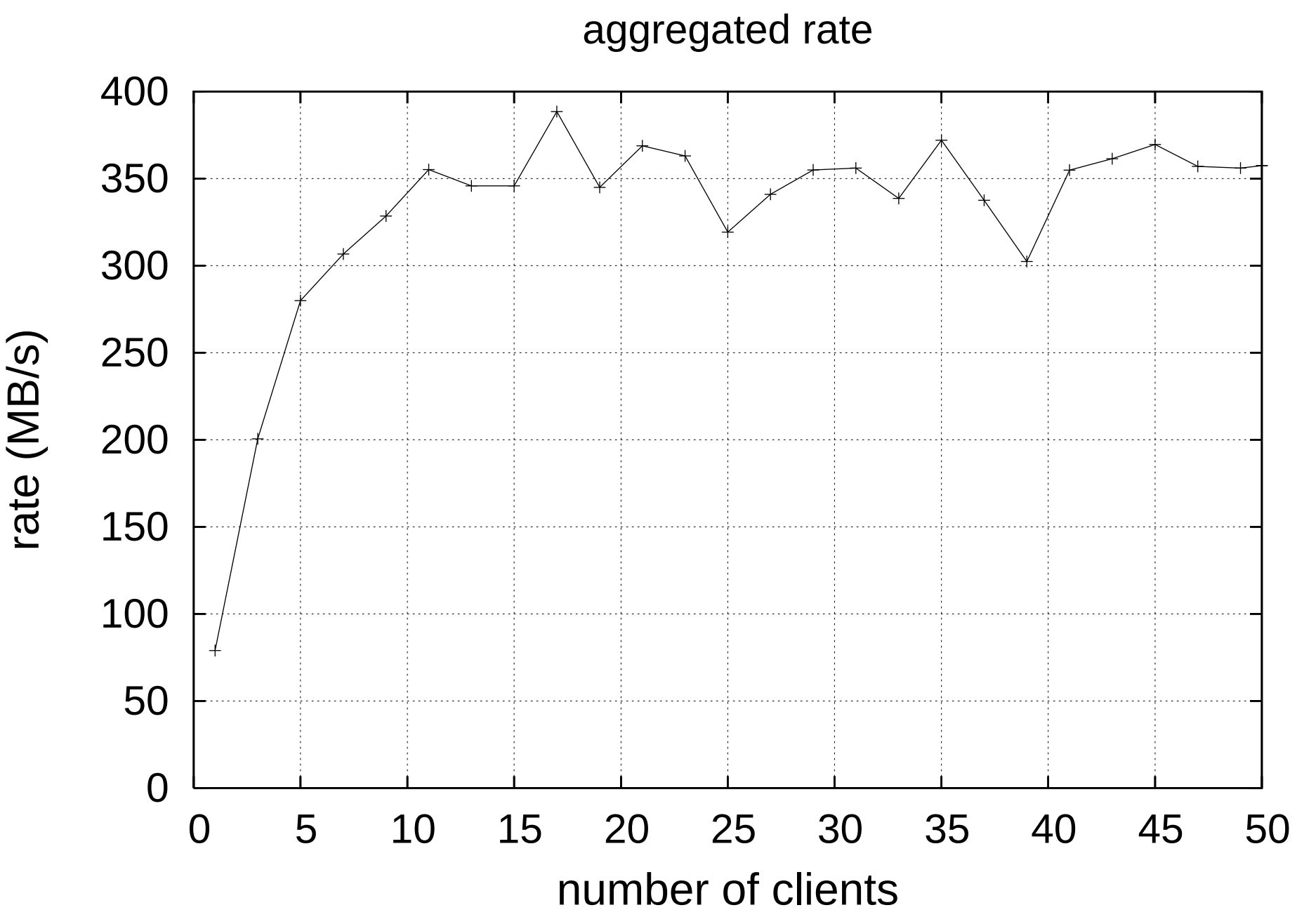

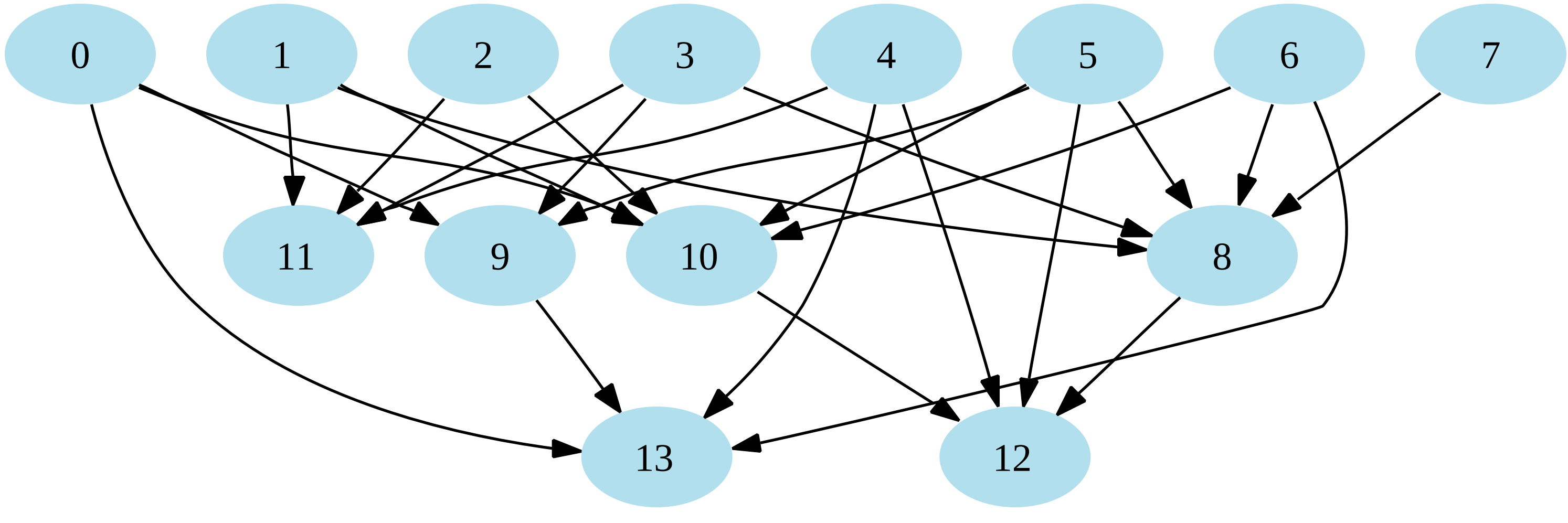

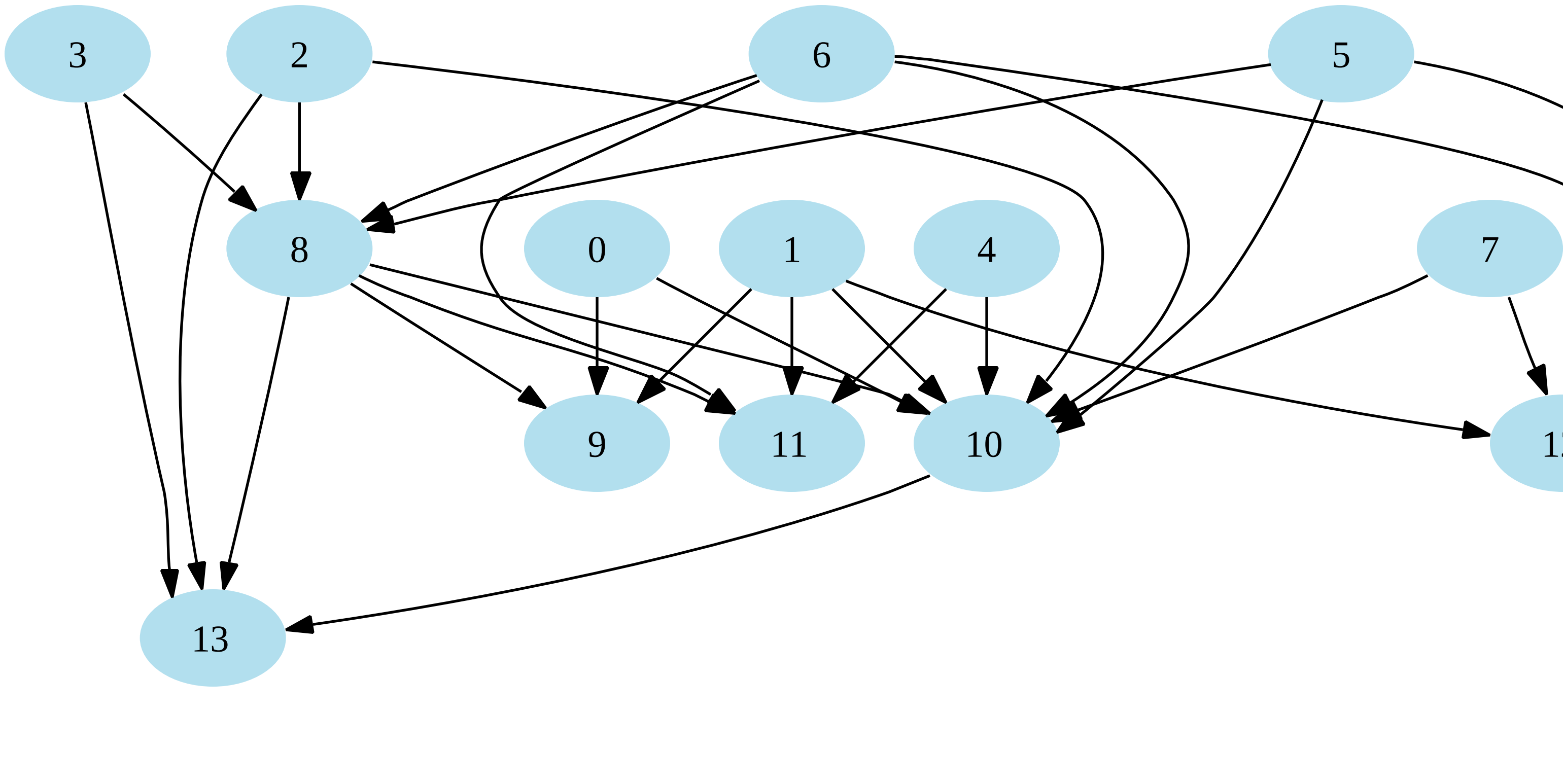

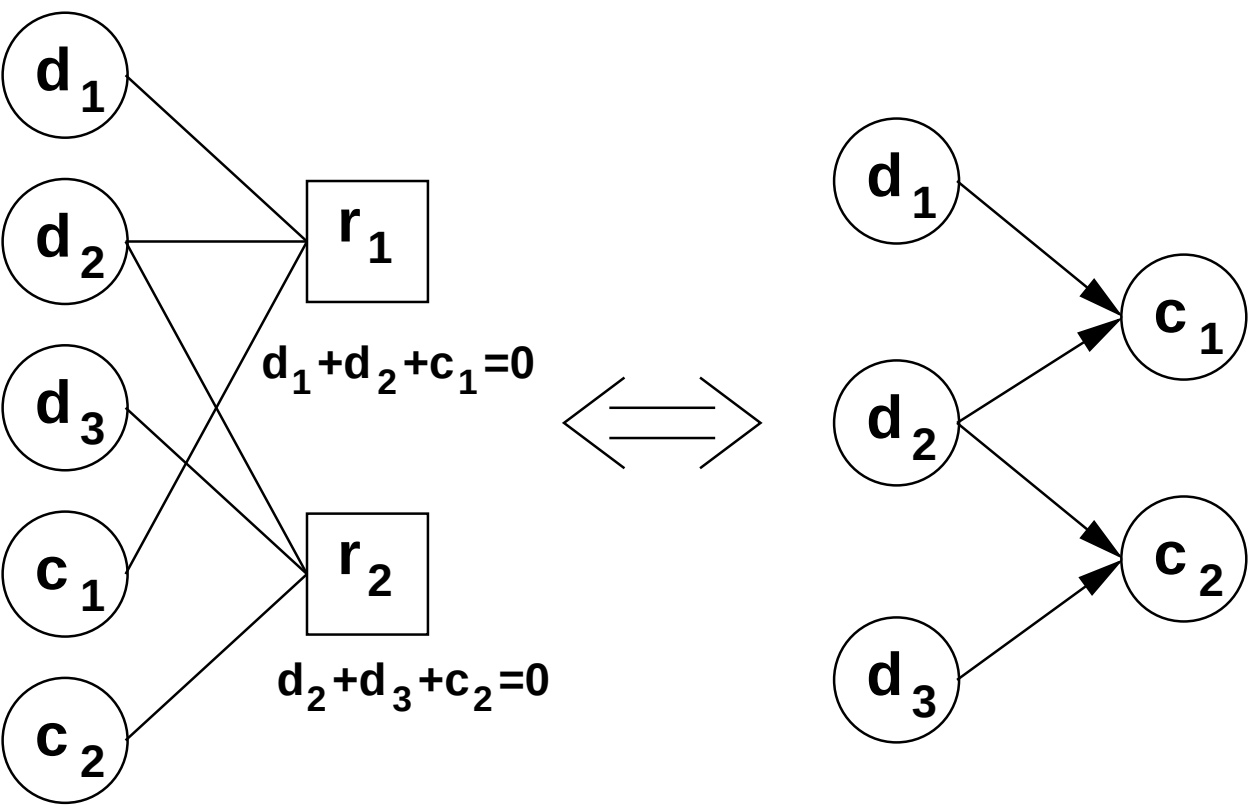

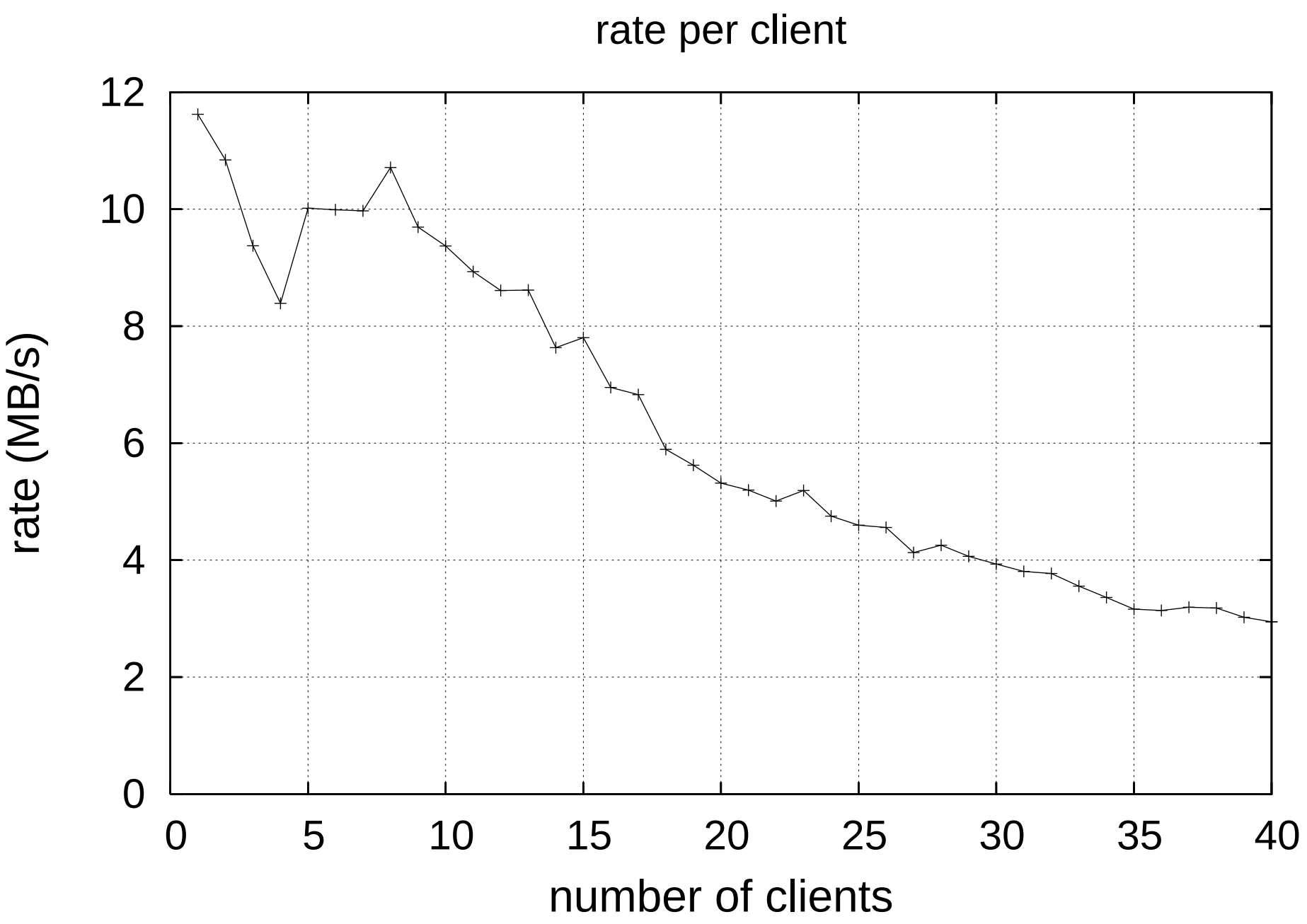